\documentclass[11pt]{article}
\usepackage{moriond,epsfig}
\usepackage{lineno}
\usepackage{amssymb} 
\usepackage{amsmath} 
\usepackage[utf8]{inputenc} 

\def\Journal#1#2#3#4{{#1} {\bf #2}, #3 (#4)}

\def\PLB{{\em Phys. Lett.}  B}
\def\PRL{\em Phys. Rev. Lett.}

\def\be{\begin{equation}}
\def\ee{\end{equation}}
\def\bea{\begin{eqnarray}}
\def\eea{\end{eqnarray}}

\newcommand{\infb}{fb$^{-1}$}
\usepackage{multirow}

\newcommand{\ptt}{\ensuremath{p_{{\mathrm T_t}}}}
\newcommand{\hgg}{\ensuremath{H \rightarrow \gamma\gamma}}

\newcommand{\hWWlnln}{\ensuremath{H \rightarrow WW^{(*)} \rightarrow \ell^+\nu\ell^-\overline{\nu}}}

\newcommand{\hWWlnqq}{\ensuremath{H \rightarrow WW \rightarrow \ell\nu q\overline{q}'}}

\newcommand{\hZZllnn}{\ensuremath{H \rightarrow ZZ\rightarrow \ell^+\ell^-\nu\overline{\nu}}}
\newcommand{\hZZllqq}{\ensuremath{H \rightarrow ZZ\rightarrow \ell^+\ell^- q\overline{q}}}
\newcommand{\hZZllll}{\ensuremath{H \rightarrow ZZ^{(*)}\rightarrow \ell^+\ell^-\ell^+\ell^-}}
\newcommand{\htt}{\ensuremath{H \rightarrow \tau^+\tau^-}}

\newcommand{\mh}{\ensuremath{m_{H}}}

\newcommand{\ZZbkg}{\ensuremath{ZZ^{(*)}}}

\newcommand{\hWlvbb}{\ensuremath{WH \rightarrow \ell\nu b\overline{b}}}
\newcommand{\hZllbb}{\ensuremath{ZH \rightarrow \ell^+\ell^- b\overline{b}}}
\newcommand{\hZvvbb}{\ensuremath{ ZH \rightarrow \nu \overline{\nu} b\overline{b} }}

\newcommand{\httll}{\ensuremath{H\rightarrow \tau\tau \rightarrow \ell^+\ell^-4\nu }}
\newcommand{\httlh}{\ensuremath{H\rightarrow \tau\tau \rightarrow \ell \tau_{had} 3 \nu }}
\newcommand{\htthh}{\ensuremath{H\rightarrow \tau\tau \rightarrow \tau_{had} \tau_{had} 2 \nu }}
\newcommand{\htthhj}{\ensuremath{H\rightarrow \tau\tau \rightarrow \tau_{had} \tau_{had} 2 \nu + {\rm jet}}}

\begin{document}
\vspace*{4cm}
\title{Searches for the Standard Model Higgs Boson with the ATLAS Detector}

\author{Ralf Bernhard}

\address{on behalf of the ATLAS collaboration\\Physikalisches Institut, %
Albert-Ludwigs Universit\"at Freiburg, Germany}

\maketitle\abstracts{
The most recent results for searches for the Standard Model Higgs boson at a center-of-mass energy of $\sqrt{s}=$7 TeV using 4.9~\infb\ of data collected with the ATLAS detector at CERNs Large Hadron Collider are presented. 
}
\section{Introduction}\label{sec:introduction}
In the Standard Model (SM) of particle physics the Higgs mechanism is responsible for breaking electroweak symmetry, thereby giving mass
to the $W$ and $Z$ bosons. It predicts the existence of a heavy scalar boson, the Higgs boson, with a mass that can not be predicted by the SM.
Direct searches for the Higgs Boson were performed at the LEP experiments and yielded a direct mass limit of $\mh$ $>$ 114.4 GeV~\cite{LEPlimit} and at the Tevatron 
excluding the region $156 < \mh < 177$ GeV~\cite{tevWinter2012}. 
Indirect limits have been placed on the Higgs boson mass by the LEP, SLD and Tevatron 
experiments from electroweak precision measurements \cite{EWconst}.
The SM fit yields a best value of $m_H=94^{+29}_{-24}$~\cite{LEPfit}. The corresponding upper limit on the Higgs mass at $95\%$ CL is $m_H<152$~GeV. 
\section{Search Channels}
In contrast to the combination of searches presented in [5] all analyses now use the full dataset of 4.9~\infb\ recorded in 2011, as shown in Tab.~\ref{tab:channels} which also
indicates the mass range of the analysis.
To enhance the sensitivity, the analysis channels under study are divided into sub-channels with different signal to background ratios
or with a different sensitivity to various systematic uncertainties. 
In the following the search channels are described.
\begin{table}[htb]
\tiny
\vspace{1em}
\begin{center}
\begin{tabular}{c|c|c|c|c}\hline\hline
\multirow{2}{*}{Higgs Decay channel}      & \multirow{2}{*}{Additional Sub-Channels}         &$\mh$   &  \multirow{2}{*}{L [\infb]} & Ref. \\ 
                                                    &              & Range [GeV]&    \\ \hline\hline
low-$\mh$, good mass resolution  &  & & \\ \hline
\multirow{1}{*}{$H\to\gamma\gamma$}   & 9 sub-channels (\ptt $\otimes \eta_\gamma \otimes \rm conversion$) & 110-150 & 4.9  &[6] \\
\multirow{1}{*}{$H\to ZZ \to \ell\ell\ell'\ell'$}  & $\{4e,2e2\mu,2\mu2e,4 \mu\}$  & 110-600 & 4.8   &[7] \\ \hline
low-$\mh$, limited mass resolution  & & \\ \hline
\multirow{1}{*}{$H\to WW \to \ell\nu\ell\nu$} & $\{ee,e\mu,\mu\mu\}$ $\otimes$ \{0-jet, 1-jet, VBF\}  & 110-300-600 & 4.7   &[8]  \\
\multirow{3}{*}{$VH\to b\overline{b}$} &  $Z\to\nu \overline{\nu}$ & \multirow{3}{*}{110-130} & \multirow{3}{*}{4.6}  & \multirow{3}{*}{[9]}\\
& $W\to\ell\nu$ && \\ 
& $Z\to \ell\ell$ && \\ 
\multirow{1}{*}{$H\to \tau^+\tau^- \to \ell\ell4\nu$} & $\{e\mu\}\otimes \{$0-jet$\}$  $\oplus$ \{1-jet, VBF, $VH\}$ & 110-150 & 4.7   &[10]\\
\multirow{2}{*}{$H\to \tau^+\tau^- \to \ell\tau_{\rm had} 3\nu$} & $\{e,\mu \}$ $\otimes$ \{0-jet\} $\otimes$  $\{ E_T^{\rm miss}  \gtrless 20~\textrm{GeV}\} $ & \multirow{2}{*}{110-150} & \multirow{2}{*}{4.7}  & \multirow{2}{*}{[10]} \\
 & $\oplus$  $\{e,\mu \}$ $\otimes$ \{1-jet, VBF\}  &  &    \\
\multirow{1}{*}{$H\to \tau^+\tau^- \to \tau_{\rm had}\tau_{\rm had} 2\nu$} & \{1-jet\} & 110-150 & 4.7  & [10] \\ \hline
high-$\mh$   & & \\ \hline
\multirow{1}{*}{$H\to ZZ \to \ell\ell\nu\bar{\nu}$}  & $\{ee,\mu\mu\}$ $\otimes$ \{low pile-up, high pile-up\} & 200-280-600 & 4.7  & [11] \\
\multirow{1}{*}{$H\to ZZ \to\ell\ell q\bar{q}$} &  \{$b$-tagged, untagged\}      & 200-300-600 & 4.7  & [12] \\ 
\multirow{1}{*}{$H\to WW \to \ell\nu q\overline{q'}$}  & $\{e,\mu\}$ $\otimes$ \{0-jet, 1-jet\}  & 300-600 & 4.7  & [13] \\ \hline\hline
\end{tabular}
\caption{Summary of the individual channels under study in ATLAS and contributing to the combination. 
~\label{tab:channels}}
\end{center}
\end{table}

\subsection{\hgg}
\vspace{-0.5em}
Despite the low branching ratio ($\approx$0.2\%) the diphoton decay mode is one of the most important channels in the search for the SM Higgs boson in the low 
mass region. The analysis separates events into nine independent categories based on the pseudo-rapidity of the photons, whether it was reconstructed as a
converted or unconverted photon, and on the momentum component of the diphoton system transverse to the thrust axis (\ptt)
The diphoton invariant mass $m_{\gamma\gamma}$ is used as a discriminating variable to distinguish signal and background, to take advantage of the mass 
resolution of approximately 1.4\% for $\mh$ around 120 GeV. The distribution of $m_{\gamma\gamma}$ in the data is fit to a smooth function to estimate the background.
Figure~\ref{fig-hgg} (left) shows the inclusive invariant mass distribution of the observed candidates, summing over all categories.
\subsection{\hZZllll}
\begin{figure}[htb] 
\begin{center} 
\includegraphics[width=0.54\textwidth]{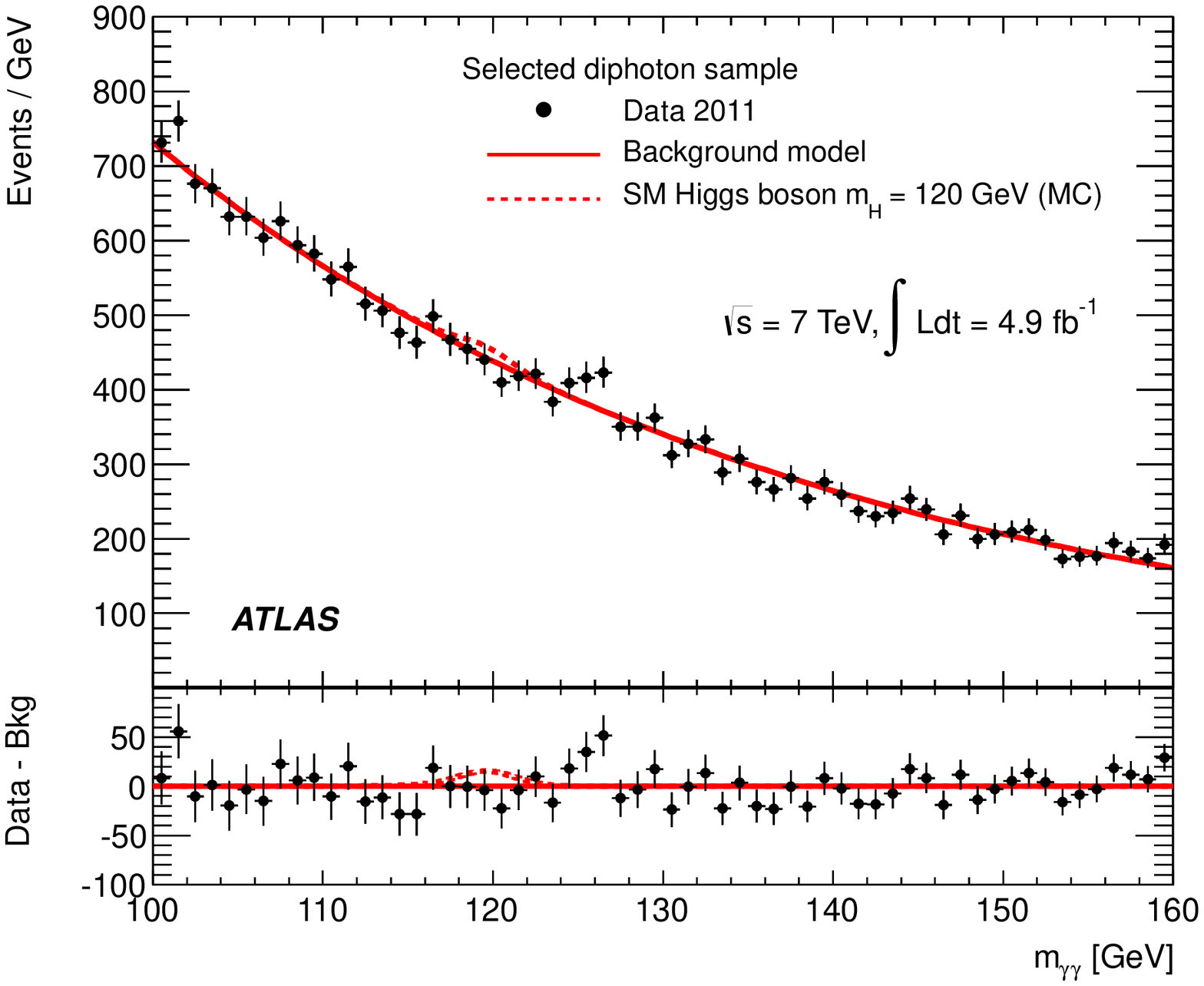}
\includegraphics[width=0.44\textwidth]{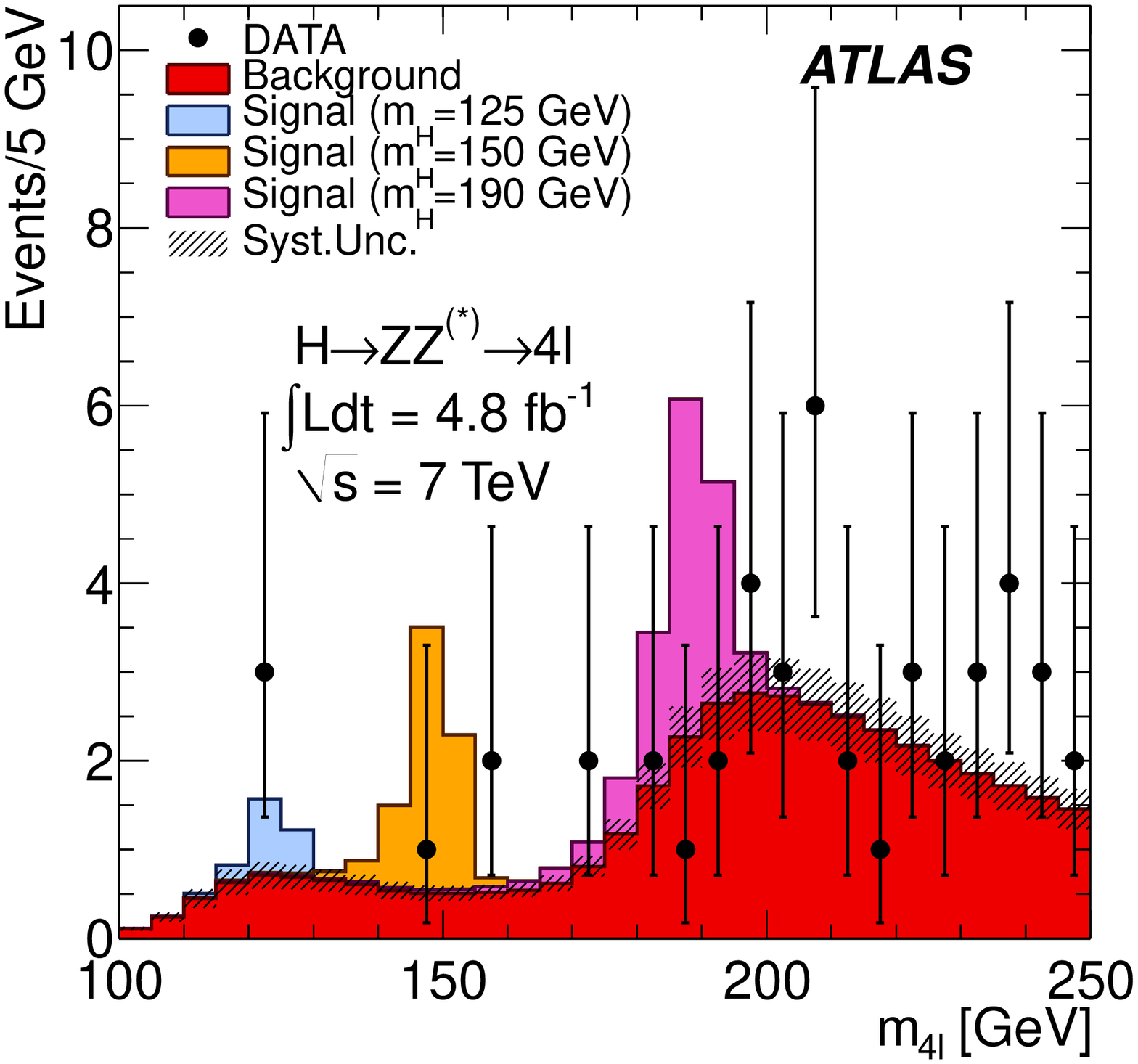}
\end{center}
\caption[]{Distributions of the reconstructed invariant mass for the selected candidate events and for the total background and signal expected in the \hgg\ (left) and the 
\hZZllll\ (right).}
\label{fig-hgg} 
\end{figure}

\vspace{-0.5em}
In this search the events are categorised according to the lepton flavour combinations.
The main irreducible \ZZbkg\ background is estimated using Monte Carlo simulation and the reducible $Z$+jets is estimated from control regions in the data. 
The mass resolutions are approximately 1.5\%\ in the four-muon channel and 2\%\ in the four-electron channel for
\mh$\sim$120~GeV. The four-lepton invariant mass is used as a discriminating variable and its distribution for events 
selected after all cuts shown in Fig.\ref{fig-hgg} on the right side.
\subsection{\hWWlnln}
\begin{figure}[htb] 
\begin{center} 
\includegraphics[width=0.48\textwidth]{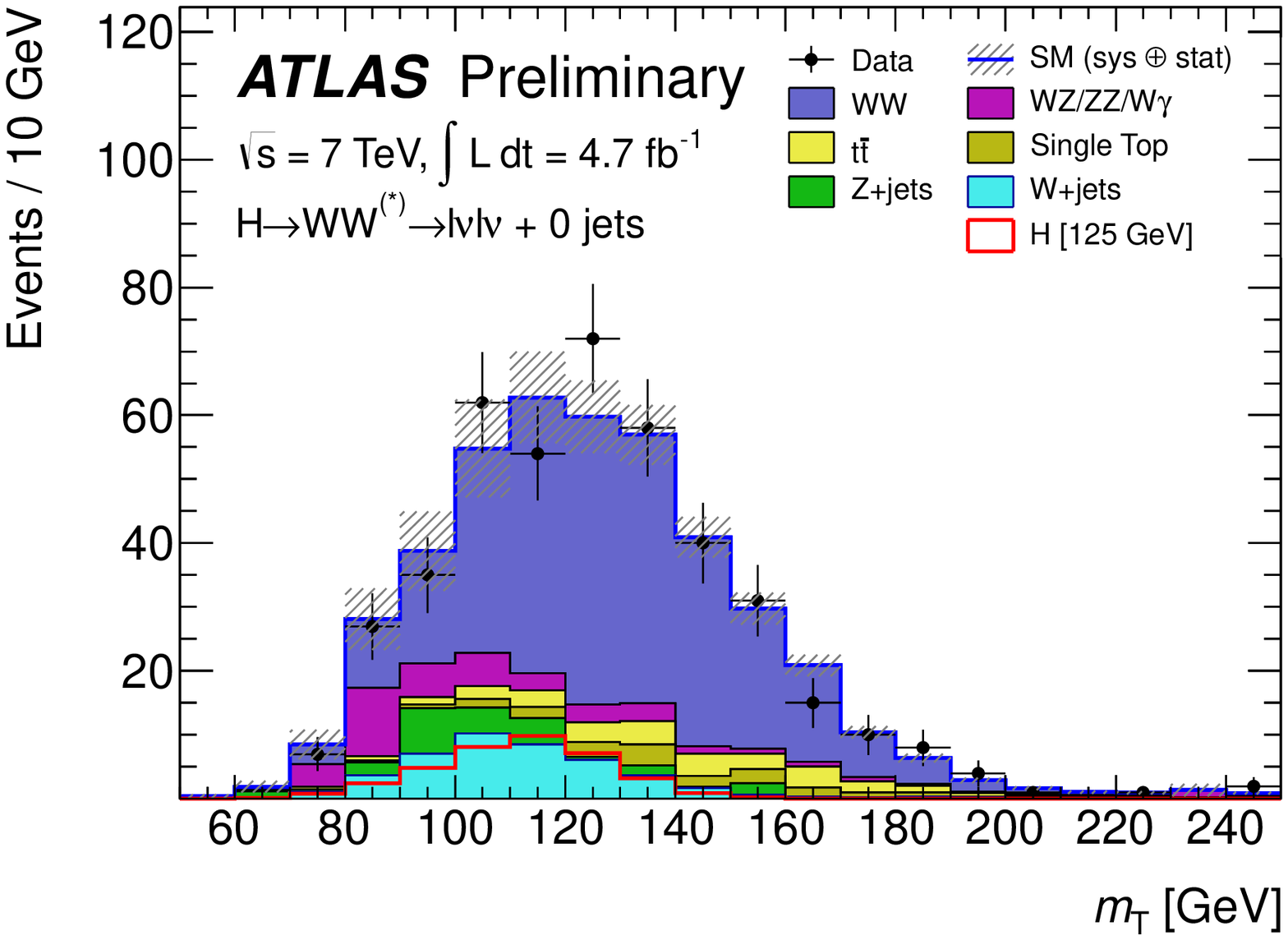}
\includegraphics[width=0.48\textwidth]{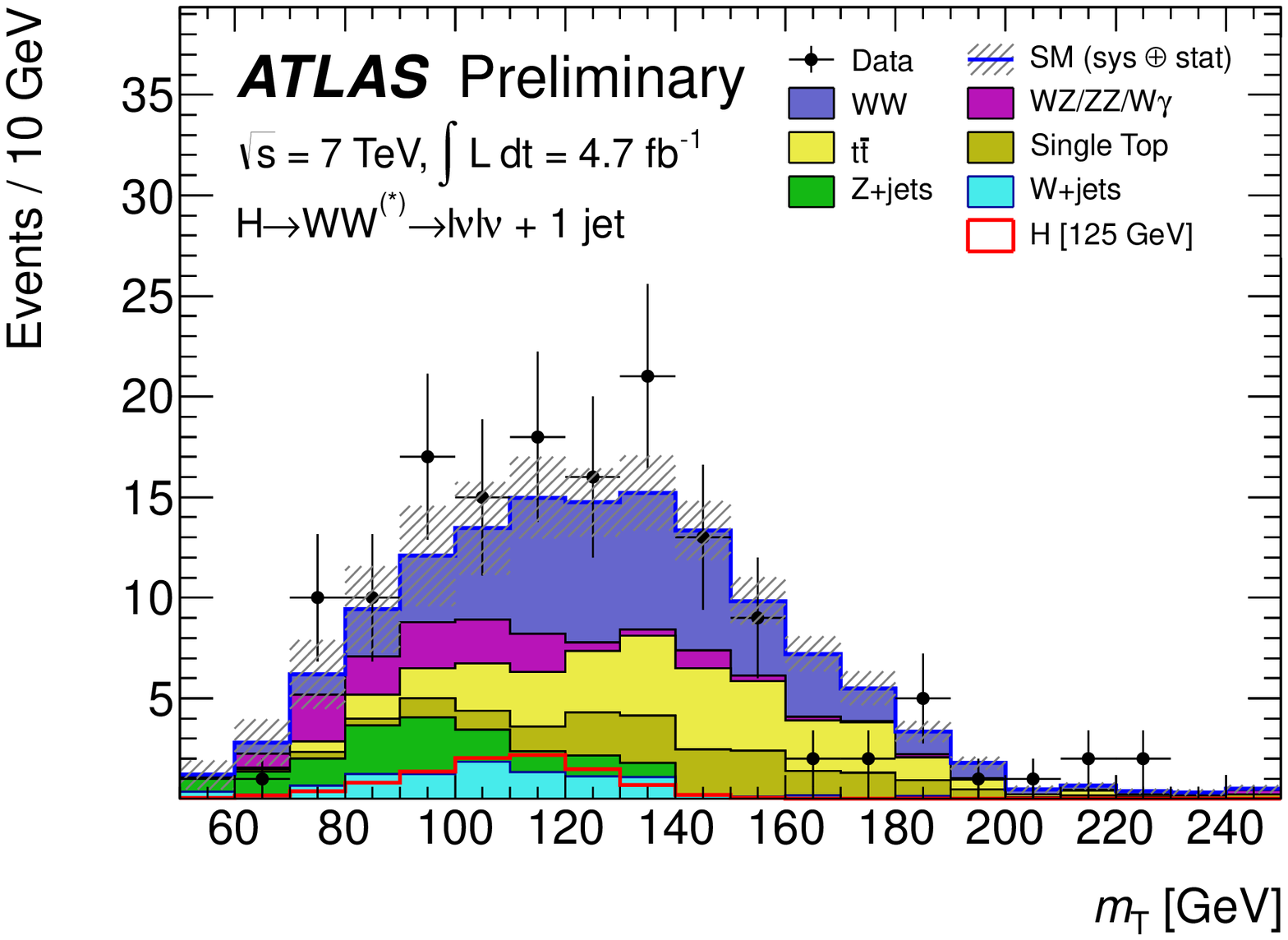}
\end{center}
\caption[]{Distributions of the reconstructed transverse mass for the selected candidate events and for the total background and signal expected in the \hWWlnln\ channel for events with 0-jets (left) and1-jets (left).
}
\label{fig-ww} 
\end{figure}

\vspace{-0.5em}
The analysis is separated into 0-jet, 1-jet and 2-jet categories as well as according to lepton flavour. 
The main backgrounds are estimated from the data using control regions and extrapolating into the signal region using Monte Carlo simulation.
As a discriminating variable the $WW$ transverse mass ($m_T$) distribution is used, which is shown 
for events with 0-jets  and 1-jets in Fig.~\ref{fig-ww} on the left and right side, respectively.  

\subsection{\hZllbb, \hWlvbb, \hZvvbb}
\vspace{-0.5em}
All three analyses require exactly two $b$-tagged jets and the invariant mass of the two $b$-jets, $m_{bb}$, is used as a discriminating
variable. To increase the sensitivity of the search, the $m_{bb}$ distribution is examined in sub-channels with different signal-to-background ratios. In the
searches with one or two charged leptons, the division is made according to four bins in transverse momentum of the
reconstructed vector boson. The individual channels are not broken into distinct lepton-flavour categories.

\subsection{\httll, \httlh, \htthh}
\vspace{-0.5em}
In the \htt\ channel any combination of events with leptonic decaying taus or hadronic decaying taus are considered. 
For the \htthhj\ channel and \httll\ channel as a discriminating variable the $\tau\tau$ invariant mass is used and estimated using the collinear approximation.
As a discriminating variable in the \httlh\ channel a Missing Mass Calculator technique is used to estimate the ditau invariant mass which does not assume a strict collinearity between the visible and invisible decay products of the tau leptons. 

\subsection{\hZZllnn,  \hZZllqq,  \hWWlnqq}
\vspace{-0.5em}
The \hZZllnn\ is split into two lepton flavour categories and analysed in the mass range from 200 to 600 GeV 
and is sensitive to a SM Higgs boson in the range of $260 \le \mh \le 490$ GeV. The \hZZllqq\ analysis 
is divided into events where the two jets are $b$-tagged and
into events with less than two $b$-tags. This analysis is expected to exclude a SM Higgs boson in the range of $360 \le \mh \le 400$ GeV at the 
95\% CL. In the \hWWlnqq\ channel the $\ell \nu q\overline{q}'$ mass distribution is used as a
discriminating variable imposing mass constraints on both W bosons. The analysis reaches the best sensitivity of two times the SM Higgs boson cross section around \mh = 400 GeV.

\section{Combination}
\begin{figure}[htb] 
\begin{center} 
\includegraphics[width=0.52\textwidth]{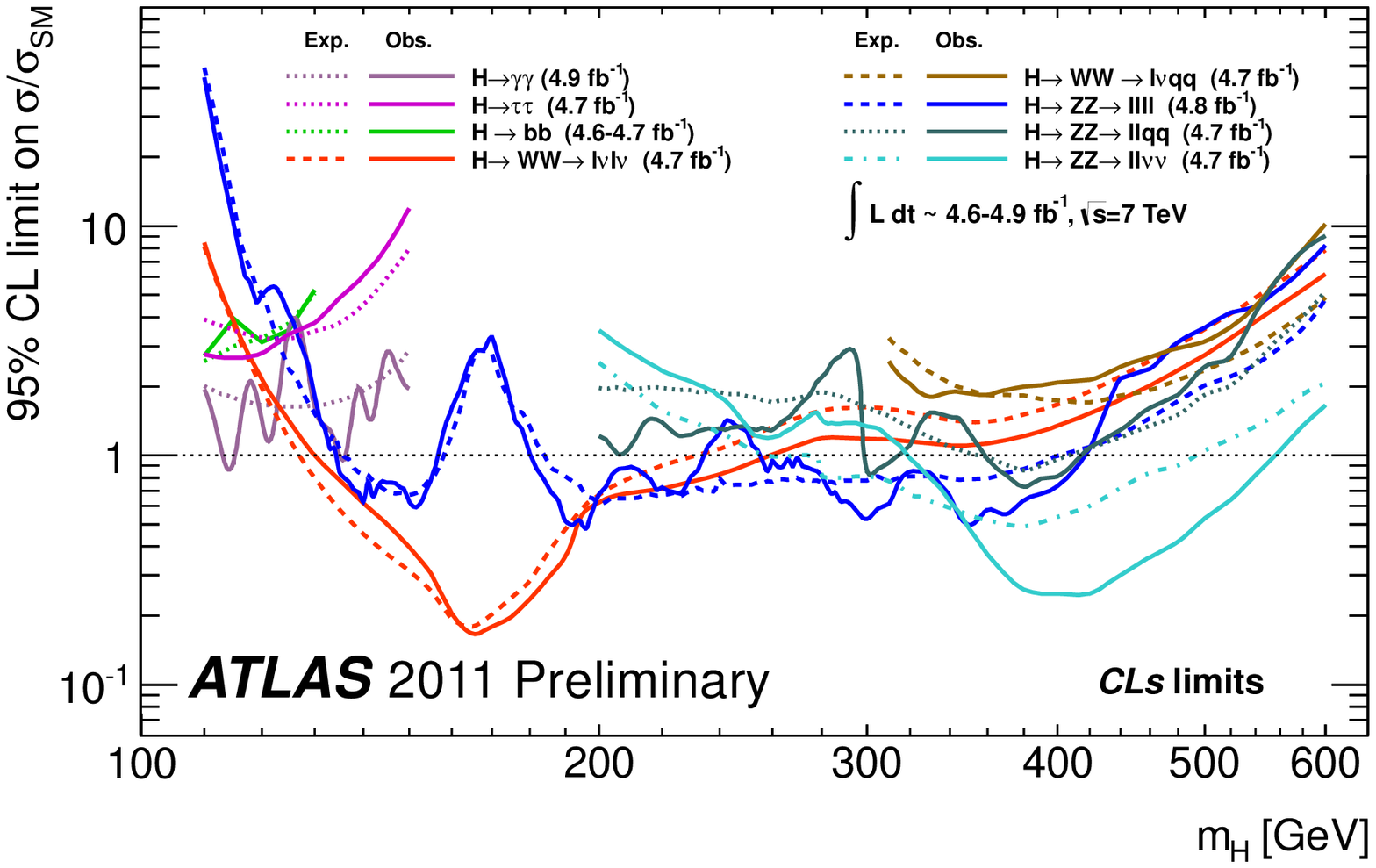}
\includegraphics[width=0.42\textwidth]{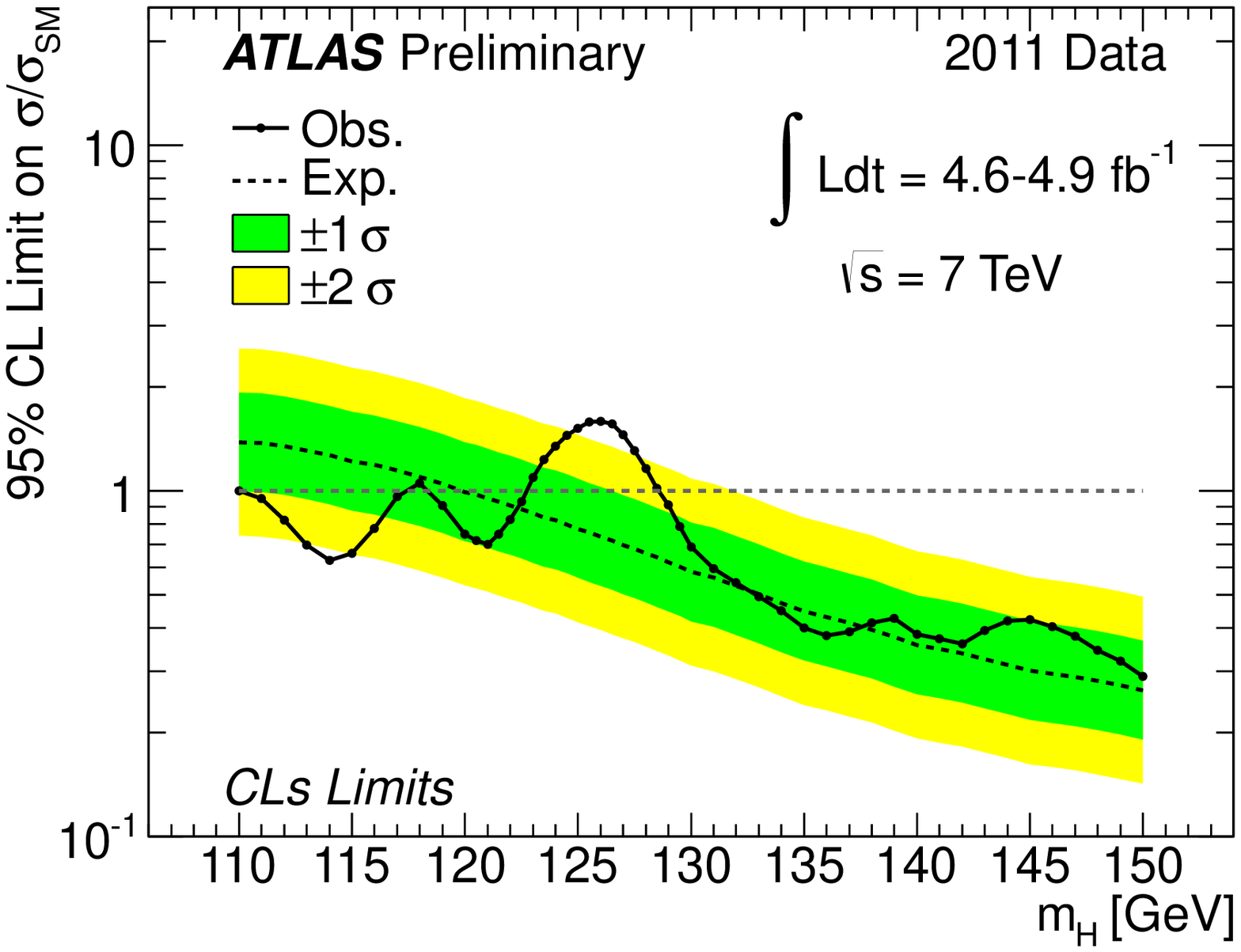}
\end{center} 
\caption[]{Left: The observed (solid) and expected (dashed) 95\%~CL cross
  section upper limits for the individual search channels as a function of the
  Higgs boson mass. Right: The observed and expected 95\%~CL
    combined upper limits on the SM Higgs boson production cross as a function of
    \mh\ in the low mass range. 
}
\label{fig-limits} 
\end{figure}

\vspace{-0.5em}
The combination procedure is based on the profile likelihood ratio test statistic.
The signal strength, $\mu$, is defined as the ratio of a given Higgs
boson production cross section ($\sigma$) to its SM value ($\sigma_{SM}$), $\mu=\sigma/\sigma_{\rm SM}$.
Exclusion limits are based on the $CL_s$ prescription~\cite{Read:2002hq}; a value of $\mu$ is regarded
as excluded at the 95\% (99\%)~CL when $CL_s$ takes on the corresponding value.
Figure~\ref{fig-limits} (left) shows the expected and observed limits from the individual channels as described above entering the combination.
For the low mass region (below \mh $< 150$ GeV) the combined 95\%~CL exclusion limits~\cite{higgsComb2012} on $\mu$ 
are shown in Fig.~\ref{fig-limits} (right) as a function of \mh.  
An excess of events is observed near \mh$\sim$126~GeV in the \hgg\ and \hZZllll\ channels, both of which provide fully reconstructed 
candidates with high-resolution in invariant mass. 
\section{Conclusions}
The full dataset recorded in 2011 by the ATLAS experiment has been used to update searches for the SM Higgs boson. 
A Higgs boson with a mass in the ranges from 110.0 GeV to 117.5 GeV, 118.4 GeV to 122.7 GeV, and 128.6 GeV to 529.3 GeV  is
excluded at the 95\% CL, while in the absence of a signal the range 119.8 GeV to 567 GeV is expected to be excluded. Between 130~GeV and 486~GeV the 
exclusion is even at the 99\%~CL. Around $m_H$ of 126 GeV an excess of events is observed with a local significance of 2.5$\sigma$. The expected significance in the presence of a SM Higgs boson at that mass hypothesis is 2.8$\sigma$.
The global probability for such an excess to occur anywhere in the explored Higgs boson mass region is estimated to be approximately 30\%,
in the range not excluded at the 99\%\ CL, it amounts to approximately 10\%.


\section*{References}
\begin{thebibliography}{99}

\bibitem{LEPlimit}R. Barate {\it et al}, 
 \Journal{\PLB}{565}{61}{2003}.

\bibitem{tevWinter2012} CDF and D0 Collaborations, arXiv:1203.3774.

\bibitem{EWconst}Tevatron Electroweak Working Group, August 2009, {\texttt{http://tevewwg.fnal.gov/}}.

\bibitem{LEPfit}LEP Electroweak Working Group, March 2012, {\texttt{http://lepewwg.web.cern.ch/LEPEWWG/}}.

\bibitem{atlas11} ATLAS Collaboration, \Journal{\PLB}{710}{49}{2012}.

\bibitem{hgg2012} ATLAS Collaboration, \Journal{\PRL}{108}{111803}{2012}.

\bibitem{hzzllll2012} ATLAS Collaboration, \Journal{\PLB}{710}{383}{2012}.

\bibitem{hww2012} ATLAS Collaboration, ATLAS-CONF-2012-012, http://cdsweb.cern.ch/record/1429660.

\bibitem{hbb2012} ATLAS Collaboration, ATLAS-CONF-2012-015, http://cdsweb.cern.ch/record/1429664.

\bibitem{htautau2012} ATLAS Collaboration, ATLAS-CONF-2012-014, http://cdsweb.cern.ch/record/1429662.

\bibitem{hzzlnln2012} ATLAS Collaboration, ATLAS-CONF-2012-016, http://cdsweb.cern.ch/record/1429665.

\bibitem{hzzllqq2012} ATLAS Collaboration, ATLAS-CONF-2012-017, http://cdsweb.cern.ch/record/1429666.

\bibitem{hwwlnuqq2012} ATLAS Collaboration, ATLAS-CONF-2012-018, http://cdsweb.cern.ch/record/1429667.

\bibitem{Read:2002hq} A.~L. Read  J. Phys. {\bf G28} (2002)  2693--2704.

\bibitem{higgsComb2012} ATLAS Collaboration, ATLAS-CONF-2012-019, http://cdsweb.cern.ch/record/1430033.
  
\end{thebibliography}
\end{document}